\documentclass[twocolumn,amsmath,amssymb,showkeys]{revtex4}

\usepackage{graphicx}
\usepackage{dcolumn}
\usepackage{bm}
\usepackage{color}
\usepackage{hhline}

\renewcommand{\ao}{a_{0}}
\newcommand{\aoi}{a_{0,i}}
\renewcommand{\ap}{{A_{\text{p}}}}
\newcommand{\api}{A_{\text{p},i}}

\newcommand{\ac}    {A_{\text{c}}}
\newcommand{\bo}    {b_{0}}
\newcommand{\boi}    {b_{0,i}}

\newcommand{\co}    {c_{0}}
\newcommand{\coi}    {c_{0,i}}

\newcommand{\dg}    {\Delta G}
\newcommand{\eqn}[1]{(\ref{#1})}
\newcommand{\Eqn}[1]{Eq.~(\ref{#1})}
\newcommand{\fig}[1]{Fig.~\ref{#1}}
\newcommand{\tab}[1]{Table~\ref{#1}}
\newcommand{\gcap}  {{g_{\text{cap}}}}

\newcommand{\gp}    {{g_{\text{p}}}}
\newcommand{\hc}    {{H_{\text{c}}}}

\newcommand{\kbt}   {k_\text{B}T}

\newcommand{\kc}    {{k_{\text{c}}}}
\newcommand{\kci}    {k_{\text{c},i}}
\newcommand{\kcbar}    {\overline{{k_{\text{c}}}}}
\newcommand{\kcbari}    {\overline{{k_{\text{c},i}}}}

\newcommand{\mupi}   {\mu_{\text{p},i}}
\newcommand{\mupj}   {\mu_{\text{p},j}}
\newcommand{\ncap}  {N_{\text{cap}}}
\newcommand{\np}  {N_{\text{p}}}

\newcommand{\rc}    {R_{\text{c}}}
\newcommand{\rcopt}   {R_{\text{c,opt}}}
\newcommand{\rld}    {{R_{\text{LD}}}}

\newcommand{\rp}    {{R_{\text{p}}}}
\newcommand{\rpi}    {R_{\text{p},i}}
\newcommand{\rr}    {\bm{r}}

\newcommand{\rw}    {{R_{\text{w}}}}
\newcommand{\rwi}    {R_{\text{w},i}}
\newcommand{\vcap}  {V_{\text{cap}}}

\newcommand{\xpi}   {x_{\text{p},i}}

\definecolor{lgray}{gray}{.75}

\newcommand{\lb}   {\left(}
\newcommand{\Lb}   {\left[}
\newcommand{\LB}   {\left\{}
\newcommand{\rb}   {\right)}
\newcommand{\Rb}   {\right]}
\newcommand{\RB}   {\right\}}

\newcommand{\rmd}{\text{d}}
\newcommand{\ii}{_{,i}}

\begin{document}

\title{Phospholipid demixing and the birth of a lipid droplet
\footnotetext{Abbreviations: 
ER, endoplasmic reticulum; 
FA, fatty acid; 
LD, lipid droplet;  
LPC, lysophosphatidylcholine; 
NL, neutral lipid; 
PA, phosphatidic acid;
PC, phosphatidyl choline;
PE, phosphatidylethanolamine;
PI, phosphatidyl inositol;
PL, phospholipid; 
PS, phosphatidyl serine;
SE, steryl ester; 
TAG, triacylglycerol;}
}

\author{J. Zanghellini}
\email{juergen.zanghellini@uni-graz.at}
\affiliation{Institute of Chemistry, University of Graz, Heinrichstra{\ss}e 28, A-8010 Graz, Austria, EU}
\author{F. Wodlei}
\affiliation{Institute of Chemistry, University of Graz, Heinrichstra{\ss}e 28, A-8010 Graz, Austria, EU}
\author{H. H. von Gr\"unberg}
\affiliation{Institute of Chemistry, University of Graz, Heinrichstra{\ss}e 28, A-8010 Graz, Austria, EU}

\begin{abstract}
The biogenesis of lipid droplets (LD) in the yeast \emph{Saccharomyces cerevisiae} was theoretically investigated on basis of a biophysical model. In accordance with the prevailing model of LD formation, we assumed that neutral lipids oil-out between the membrane leaflets of the endoplasmic reticulum (ER), resulting in LD that bud-off when a critical size is reached.

Mathematically, LD were modeled as spherical protuberances in an otherwise planar ER membrane. We estimated the local phospholipid composition, and calculated the change in elastic free energy of the membrane caused by nascent LD. Based on this model calculation, we found a gradual demixing of lipids in the membrane leaflet that goes along with an increase in surface curvature at the site of LD formation. During demixing, the phospholipid monolayer was able to gain energy during LD growth, which suggested that the formation of curved interfaces was supported by or even driven by lipid demixing. In addition, we show that demixing is thermodynamically necessary as LD cannot bud-off otherwise.

In the case of \emph{Saccharomyces cerevisiae} our model predicts a LD bud-off diameter of about $13$ nm. This diameter is far below the experimentally determined size of typical yeast LD. Thus, we concluded that if the standard model of LD formation is valid, LD biogenesis is a two step process. Small LD are produced from the ER, which subsequently ripe within the cytosol through a series of fusions.
\end{abstract}


\maketitle

\section{Introduction}
Lipid droplets (LD) are depots for neutral lipids (NL). They exist in virtually all kind of living cells, from bacteria, to yeasts, to plants and mammals. A LD consists of a hydrophobic, NL-containing core surrounded by a phospholipid (PL) monolayer containing a small amount of proteins \cite{martin2006}. In \emph{Saccharomyces cerevisiae}, LD are mainly formed by triacylglycerols (TAG) and steryl esters (SE) in roughly equal amounts \cite{leber1994,czabany2007,czabany2008}. Conversion of free fatty acids (FA) and sterols to NL and their subsequent storage in LD is an organism's strategy to risklessly save intrinsically toxic FA and sterols for later use. If required, FA and sterols may be rapidly released from LD and used as pre-fabricated building blocks for membrane lipid synthesis as well as other complex lipids, and/or as source of chemical energy \cite{murphy1999,fujimoto2008,zanghellini2008}. Also, LD are assumed to have a function in transporting sterols to the plasma membrane \cite{czabany2007}. Indeed, it is now recognized that rather than being inert storage pools, LD are remarkably flexible, dynamic organelles \cite{fujimoto2008,murphy2009,olofsson2009}.

\begin{figure}
   \begin{center}
      \includegraphics[width=3.25in]{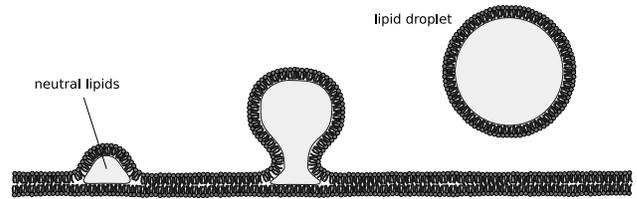}
      \caption{LD formation according to the standard model. During their biosynthesis NL ``oil-out'' in between the leaflets of the ER bilayer, forming spherical structures. Mature LD then bud-off and form independent organelles.}
      \label{fig1}
   \end{center}
\end{figure}

LD biogenesis is everything but clear. According to a widely accepted model, NL accumulate within distinct regions of the endoplasmic reticulum (ER) membrane, initially forming a lens-shaped and then a spherical bulge in the membrane (\fig{fig1}). After reaching a critical size, mature LD will bud-off, being encapsulated in a PL monolayer that is directly derived from the cytoplasmic ER leaflet \cite{murphy1999,murphy2001,martin2006,czabany2007,fujimoto2008}. This budding model is in line with several experimental findings. (i) In yeast the same ER proteins, except for very few, are also detected on the LD membrane \cite{huh2003,natter2005}. (ii) Most LD proteins lack transmembrane spanning domains \cite{athenstaedt1999}. (iii) In yeast mutants, which are unable to synthesize TAG and SE, LD do not form. Nonetheless all typical LD proteins are found in these strains, but now solely localized to the ER \cite{sorger2004} and the cytosol. It has also been hypothesized that developing LD might not bud-off, but are cut out from the ER in form of bicelles, leaving a transient hole in the ER membrane \cite{ploegh2007}.

A nascent LD trapped within the leaflets of the ER has never been observed experimentally, thus other mechanisms for its formation have been suggested. (See \cite{walther2008,thiele2008} for reviews.) The most prominent alternative is based on vesicular budding \cite{mcmahon2004,praefcke2004,corda2006}. In such a process small bilayer vesicles are formed, which are subsequently filled with NL \cite{robenek2005,robenek2006}. As yet, neither model has been conclusively verified experimentally. Whether this has to be attributed to low resolution of microscopic approaches or indicates that the proposed scenarios are wrong is still a matter of debate. In this article we take up the former position. Moreover, we here assume that LD formation takes place according to \fig{fig1}. That is to say, NL-filled bulges are formed in the cytosolic monolayer of the ER membrane, from where they subsequently bud-off. Taking this scenario seriously, we calculate biophysical consequences for the process of LD biogenesis.

Our analysis has been motivated by experimental findings in yeast that the PL monolayer composition of LD differs from the one of the ER membrane \cite{zinser1991,leber1994,tauchi-sato2002}. This suggests that the PL composition is related to the local curvature of the monolayer in agreement with similar observation  from theory and experiment \cite{roux2005,jiang2008}. Lipids with a cone-like molecular shape (inducing a positive or convex curvature) are expected to be more adapted to the spherical surface of LD than wedge-like shaped ones, which induce a negative or concave curvature. Thus, the former should be enriched on LD surfaces (relative to their value in the ER membrane) while the latter ought to be depleted. For instance, due to their geometrical shape we expect to find more lysophosphatidylcholine (LPC) and less phosphatidylethanolamine (PE) in the membrane of LD compared to the ER. 

In contrast to the well established view that PL demixing is curvature-dependent, we here argue the converse, i.e., that the demixing of a lipid membrane supports the generation of membrane curvature. We use yeast LD as an example and present calculations based on the standard model of LD formation. They reveal that due to lipid demixing, the PL monolayer is able to gain energy during its shape transition, thus supporting LD formation. Additionally, we show that depending on its volume and the curvature of its surface, a nascent LD is prevented from budding-off through a demixing-controlled energy barrier, whose height decreases with increasing LD volume. We predict that at a LD diameter of about 13 nm, this barrier completely vanishes and the LD is released from the ER. We thus address two essential questions \cite{murphy1999}: How do nascent LD bud-off from membranes?  How is the mature size of LD determined?

\section{Theoretical model}
To study local deformations in an extended lipid monolayer we use the
Helfrich Hamiltonian \cite{helfrich1973},
\begin{equation}
  \label{Helfrich}
  U= \int_{\mathcal{A}} \rmd A \LB\frac{\kc}{2}\Lb H(\rr)-2\co\Rb^2 + \kcbar K(\rr)\RB,
\end{equation}
which relates the local, total and Gaussian curvature, $H=c_1+c_2$,
and $K=c_1 c_2$, respectively, of a 2-dimensional surface,
$\mathcal{A}$, to its elastic energy, $U$. Here, $c_1$ and $c_2$
denote the two principal curvatures at any given point $\rr$ on the
surface. The thin lipid membrane is characterized by its spontaneous
curvature, $\co$, and its mean and Gaussian bending modulus, $\kc$,
and $\kcbar$, respectively.

\Eqn{Helfrich} describes a continuous membrane, without any reference
to its internal structure. However, to be able to model a mixed
membrane (i.e. a membrane consisting of various different types of
PL) we assume that \eqn{Helfrich} is valid not only for a monolayer
as a whole but also for every single PL molecule. Thus,
\begin{equation}
  \label{individualH}
  U_i (\rr)= \api \LB\frac{\kc\ii}{2}\Lb H(\rr)-2\coi\Rb^2 + \kcbar\ii
  K(\rr)\RB,
\end{equation}
denotes the potential energy of a single lipid where the index $i$ in
the following distinghuishes between the different types of PLs
(PE, LPC, $\ldots$, see below).  In this equation we have assumed that
across each molecule's headgroup area the change in the local membrane
curvature is so small, that the integration in \eqn{Helfrich}
simplifies to a multiplication with the pivotal area, $\api$, occupied
by a lipid of type $i$. The pivotal area of lipid, $\api$, is defined
as the area that remains unchanged in its size upon spherical bending
\cite{rand1990}. (See \fig{fig:lipstruc} for an illustration.) Note
that in writing \eqn{individualH} we disregard contributions due to
orientational ordering \cite{kraljiglic02}, and implicitly assume
properly aligned lipids.

We may construct the PL monolayer's average free Gibbs energy, $G$, per
lipid as the sum of independent, single molecule energies, $U_i$, plus
their corresponding configuration entropies, i.e.
\begin{equation}
  \label{gibbsPerLipid}
  g:=G/N= \frac{1}{\mathcal{A}}
\int_{\mathcal{A}} \rmd A \sum_{i=1}^n \Lb x_iU_i(\rr)+x_i\kbt\ln x_i\Rb,
\end{equation}
with, $x_i$, the local lipid mole fraction in the membrane; $n$, the
number of lipid species in the monolayer; and, $N$, the total number
of lipids. In the special case of a planar membrane ($H\rightarrow 0,
K\rightarrow 0$) $g$ approaches
\begin{equation}
  \label{gibbsPerLipidPlanar}
  \gp= \sum_{i=1}^n \xpi\mupi,\quad
  \mupi=2\kc\ii\coi^2+\kbt\ln\xpi,
\end{equation}
where we have used $\xpi$ and $\mupi$ to denote the mole fraction and
chemical potential of a flat membrane, respectively.

According to \eqn{gibbsPerLipid}, the elastic energy of a mixed
membrane depends on both its composition and the local curvature of
the surface. We are interested in whether or not the total,
interfacial energy of curved membrane surfaces can be reduced by
changing the lipid composition relative to the composition of the flat
membrane.

In equilibrium, the local lipid composition, $x_i$, in
eq.~(\ref{gibbsPerLipid}) is given by the Boltzmann factor,
\begin{equation}
  \label{molefraction}
  x_i= \frac{\exp\Lb-\lb U_i(\rr)-\mupi\rb/\kbt\Rb}
{\sum_{j=1}^n\exp\Lb-\lb U_j(\rr)-\mupj\rb/\kbt\Rb}.
\end{equation}
and can be interpreted as the probability for a lipid of type $i$ to
populate regions of energy $U_i(\rr)$, while the planar part of the
membrane acts as a reservoir, which either provides or absorbs lipids
depending on the difference between $x_i$ and $\xpi$. Note that for a
flat membrane \eqn{molefraction} recovers the limiting planar lipid
composition, i.e. $(H\rightarrow 0, K\rightarrow 0) \Rightarrow
x_i\rightarrow\xpi$.

\subsection{Shape approximation}

\Eqn{gibbsPerLipid} and \eqn{molefraction} allow to estimate the
energy change caused by local adaptations of both the lipid
composition and the surface curvature. Next, we make further
assumptions regarding the shape evolution of emerging LD that form
from initially flat ER membranes.

In every stage of their biogenesis, nascent LD are assumed to form perfectly spherical protuberances of cap radius, $\rc$. The spherical cap approximation supposes a sudden transition from the planar to the spherical region of the membrane.
\begin{figure}
   \includegraphics[width=\columnwidth]{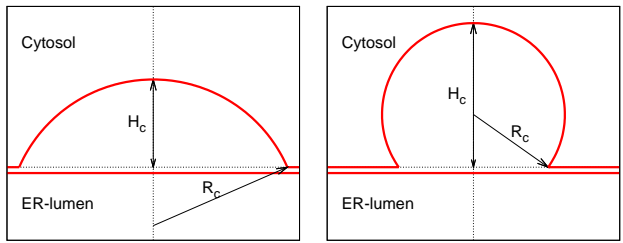}
   \caption{A LD is modeled as spherical cap of radius, $\rc$, and height $\hc$, in an otherwise planar membrane (sectional view). Thick lines represent the cytosolic and luminal leaflets of the ER membrane. The two panels illustrate situations for different values of $\rc$, but constant cap volume, $\vcap$.}
   \label{fig:ldcapneck}
\end{figure}

\fig{fig:ldcapneck} illustrates two possible geometrical configurations. In both cases the total volume of the LD, $\vcap$, is identical. In the following, it is convenient to discuss the effect of the LD volume in terms of the radius, $\rld$, of an associated, full sphere of equal volume.

Let $\rc$ denote the radius of a spherical cap, then the cap surface
$\ac(\rc,\rld)$ enclosing a cap volume $\vcap=4\pi\rld^3/3$, as well
as its corresponding pinch-off area $A_1(\rc,\rld)$, can be written as
\begin{eqnarray}
\label{he0}
A_c(\rc,\rld)\!\!&=&\!\!2\pi\rc^2 H(\rld/\rc),\\
A_1(\rc,\rld)\!\!&=&\!\!\pi \rc^2\left[2H(\rld/\rc)-H^2(\rld/\rc)\right]\!\!,\quad\\
  H(x)\!\!&=&\!\!1-2\cos\left[\frac{1}{3}\arccos(1-2x^3)+\frac{\pi}{3}\right]\!\!,
\end{eqnarray}
where $\hc=\rc H(\rld/\rc)$ gives the total height of the spherical
calotte (see \fig{fig:ldcapneck}). While $\rld$ is a measure of the LD
volume, the radius $\rc$ is the representative of the shape of the
nascent LD. A LD detaches if its pinch-off area, $A_1$, vanishes. Thus
for matured LD $\rld=\rc$ and $H(1)=2$, such that $A_1=0$ and
$\ac=4\pi R^2$.

\subsection{Free Gibbs energy of a LD}
In the spherical cap approximation, the local and Gaussian curvature
is independent of the position on the cap surface and simply given by
$H=2/\rc$ and $K=1/\rc^2$, respectively. Thus, the free Gibbs energy
per lipid,
\begin{equation}
  \label{gcap}
  \gcap(\rc)= \sum_{i=1}^n \Lb x_iU_i(\rc)+x_i\kbt\ln x_i\Rb,
\end{equation}
depends parametrically on the cap radius.

With an expression for the surface at hand it is possible to estimate the average number of lipids on nascent LD as $\ncap=\sum_{i=1}^n x_i A_c/\api$, as well as the number of lipids, which originally were found in the pinch-off area but migrated as $\np= \sum_{i=1}^n \xpi A_1/\api$. Then, the total Gibbs energy of formation for a LD becomes
\begin{equation}
 \label{eqn:totdg}
  \dg= \ncap\gcap-\np\gp.
\end{equation}

In the following we will keep the total LD volume, $\vcap$, constant and ask which configuration, out of all possible values of $\rc$, is energetically favored.

\subsection{Parameter estimation}
Glycero-PL, in particular phosphatidic acid (PA), phosphatidyl choline (PC), PE, phosphatidyl inositol (PI), LPC and phosphatidyl serine (PS), are major constituents of ER membranes \cite{zinser1991}. According to \eqn{Helfrich}, our modeling approach requires four characteristic numbers per lipid species $i$. Two of which provide structural information ($\api$ and $\coi$), while the remaining two describe elastic properties ($\kci$ and $\kcbari$). The latter were found to vary little for different lipid species \cite{marsh2007}, and experimental results suggest that
\begin{equation}
  \label{emod}
  \kcbari\approx 0.8\kci,
\end{equation}
irrespective of specific lipids \cite{marsh2007}.

However, literature values on geometrical data for lipids scatter
substantially. Partly because data, such as the head-group area or
chain-length, depend on temperature, salt content, the phase of the
membrane, etc., but also because of variations in experimental
methods. For instance, \cite{nagle2000} presents ten published
head-group areas for fluid phase Dipalmitoyl-PC, ranging from 57
\AA$^2$ to 73 \AA$^2$. We have compiled representative values of
structural data of lipids (\tab{tab1}), being fully aware that these
values may only be considered as rough estimates. A thorough
discussion of our estimations may be found in the supplementary
material \ref{sec:lipest}.

Here we describe lipids based on a geometrical packing parameter, $S_i$, which grossly characterizes the shapes of lipids by relating their entire volume $V_i$, to the volume given by their head-group areas $\aoi$, times the lipid length $l_i$, i.e. $S_i=V_i/(\aoi l_i)$. Then the spontaneous  curvature, $\coi$, of a monolayer of identical lipids for a cylindrical system  can be expressed as \cite{marsh1996,israelachvili1992}
\begin{equation}
  \label{scurv}
  \coi= \frac{1}{\rwi}=\frac{2}{l_i}\lb 1-S_i\rb=\frac{1}{l_i}\lb 1-\frac{\boi}{\aoi}\rb.
\end{equation}

\begin{figure}
   \begin{center}
      \includegraphics[width=\columnwidth]{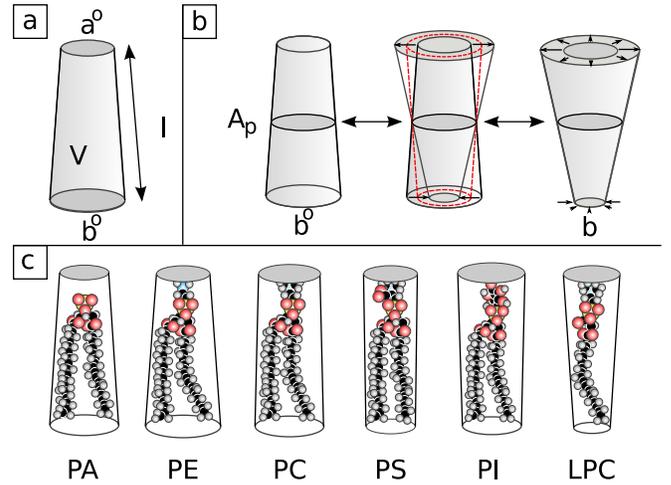}
      \caption{Characterization of different PL according to their structure factor, $S_i=(1+\boi/\aoi)/2$, panel a and c. $S>1$ indicates cone like shape, which favors concave membranes. $S<1$ describes lipids as inverted cones, which produce convex curvatures. Panel b illustrates the pivotal plane, which by definition does not change in size upon spherical bending.}
      \label{fig:lipstruc}
   \end{center}
\end{figure}

In the last part of the equation we have approximated the volume of a PL by the volume of a truncated cone, $V_i=(\aoi+\boi) l_i /2$, where $\aoi$ and $\boi$ are the lipid's head- and base area, respectively (\fig{fig:lipstruc}). $\rwi$, is typically measured in fully hydrated lipid phases and refers to the radius of curvature of the lipid-water interface. For simplicity and despite substantial variations in the lipid's length, we assume that all lipid-types are basically as long as PE, i.e. $l=l_i=22$ \AA~\cite{rand1990}. To further ease analysis we define the pivotal plane to sit midway between the base and head area of a lipid, thus
\begin{equation}
  \label{ploc}
  \rwi = \rpi + l/2.
\end{equation}
However, according to experiments the neutral plane sits closer to the boundary between the hydrocarbons and polar group. For PE a representative value is $0.37 l$ \cite{fuller2001}.

\begin{table}[b]
\caption{\label{tab1} Structural data of various PL. Literature values are marked by footnotes, estimated values are highlighted in gray, calculated values [using \eqn{emod} - \eqn{ploc}] are printed without any tag, and framed lines contain data which enter the calculation. Abbreviations: $\ao$, lipid head-group area; $\ap$, molecular area of the pivotal plane; $\bo$, base area; $S$, shape factor; $c_{0}$, spontaneous curvature of the membrane monolayer; $\rw$, cylinder radius to the lipid-solvent interface; $\rp$ cylinder radius to the pivotal plane; $\kc$ and $\kcbar$, mean and Gaussian curvature elastic modulus, respectively.}
\begin{ruledtabular}
\begin{tabular}{lllllll}
       & PE & PA & PC & PS & PI & O-LPC \\
\hline
 $\aoi$ (\AA$^2$) & 54.\footnotemark[1] & \colorbox{lgray}{45.} & 72.\footnotemark[7]\footnotemark[8] & 54.\footnotemark[11] & \colorbox{lgray}{84.} & 60.\footnotemark[1] \\\hline
\multicolumn{1}{|l}{$\api$ (\AA$^2$)} & 73.\footnotemark[1] & \colorbox{lgray}{59.} & 82.\footnotemark[9] & \colorbox{lgray}{50.} & \colorbox{lgray}{75.} & \multicolumn{1}{l|}{45.\footnotemark[1]} \\\hline
 $\boi$ (\AA$^2$) & 105.\footnotemark[1] & 73. & 87.8 & 45. & \colorbox{lgray}{65.} & 33. \medskip \\
 
  $S_i$ & 1.47 & 1.31 & 1.11 & 0.92 & 0.89 & 0.78 \\\hline
\multicolumn{1}{|l}{$\coi$ (\AA$^{-1}$)} & -0.043 & -0.029 & -0.010 & 0.0075 & 0.01 & \multicolumn{1}{l|}{0.020} \\\hline
$\rwi$ (\AA) & -23.\footnotemark[2]  & -35. & -100. & 133. & 100. & 49. \\
$\rpi$ (\AA) & -28.5\footnotemark[3] & -46.\footnotemark[6] & -87.3\footnotemark[10] & 144.\footnotemark[4] & 89. & 38.\footnotemark[1] \\
 & -30.\footnotemark[4] & & -95.\footnotemark[7] \\
 & & & -143.\footnotemark[9] \medskip \\\hline
\multicolumn{1}{|l}{$\kci$ ($\kbt$)} & 11.\footnotemark[3]\footnotemark[4]\footnotemark[5] & \colorbox{lgray}{10.} & 9.\footnotemark[10] & 10.\footnotemark[4]  & \colorbox{lgray}{10.} & \multicolumn{1}{l|}{\colorbox{lgray}{10.}} \\
\multicolumn{1}{|l}{$\kcbar\ii$ $(\kbt)$} & 8.8 & 8. & 7.2 & 8. & 8. & \multicolumn{1}{l|}{8.} \\\hline
\end{tabular}
\end{ruledtabular}
\begin{minipage}[t]{0.17\columnwidth}
\footnotetext[1]{Ref. \cite{fuller2001}}
\footnotetext[6]{Ref. \cite{kooijman2005}}
\footnotetext[11]{\mbox{unpublished, G. Pabst}}
\end{minipage}\hfill
\begin{minipage}[t]{0.17\columnwidth}
\footnotetext[2]{Ref. \cite{rand1990}}
\footnotetext[7]{Ref. \cite{marsh1996}}
\end{minipage}\hfill
\begin{minipage}[t]{0.17\columnwidth}
\footnotetext[3]{Ref. \cite{leikin1996}}
\footnotetext[8]{Ref. \cite{nagle2000}}
\end{minipage}\hfill
\begin{minipage}[t]{0.17\columnwidth}
\footnotetext[4]{Ref. \cite{fuller2003}}
\footnotetext[9]{Ref. \cite{szule2002}}
\end{minipage}\hfill
\begin{minipage}[t]{0.17\columnwidth}
\footnotetext[5]{Ref. \cite{marsh2007}}
\footnotetext[10]{Ref. \cite{chen1997}}
\end{minipage}
\end{table}

A classification according to the packing parameter, $S$, reveals three major groups; PE and PA have $S_i\gg 1$, forming wedge-shaped, inverted truncated cones, while PC as well as PS, with $S\approx 1$, are quite cylindrical in shape. Finally, PI and LPC have $S$ values between 0.7 and 0.9, corresponding to truncated cones with lipid foot areas much smaller than their head-group areas.

Finally, \tab{tab2} lists PL composition of membranes for both, the LD \cite{zinser1991,leber1994} and the ER \cite{zinser1991}. The two data sets agree well for PC and PE, showing a considerable reduction of both lipids in the LD membrane relative to the ER. Also, both data sets reflect a substantial increase in PA and PI. Conflicting results are obtained for PS; while the fraction of PS is doubled in \cite{zinser1991}, it is found to be slightly reduced in \cite{leber1994}. Precise values for LPC are not known, but it has been reported that LPC is enriched in the LD membrane \cite{tauchisato2002}. To study the packing effects even on LPC, we assume its $\xpi$ equal to that of PA.
\begin{figure}[t]
   \begin{center}
      \includegraphics[width=\columnwidth]{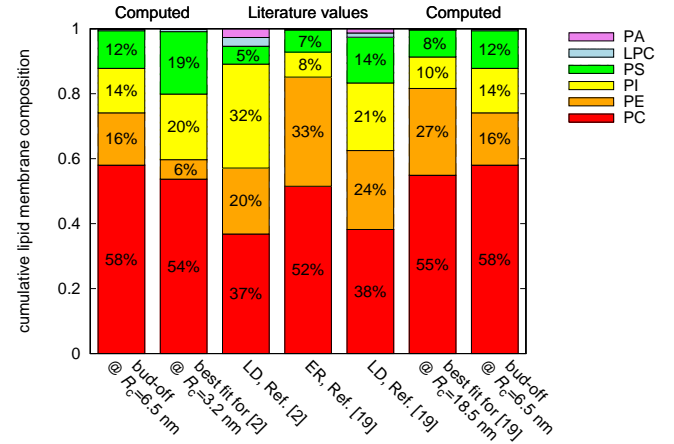}
      \caption{Literature values for experimentally measured PL composition of ER and LD membranes compared to our thermodynamic model at various radii $\rc$. Data listed in \cite{zinser1991,leber1994} for ``other lipids'' have been redistributed such that each column sums up to one. (For actual numerical values see supplemental material, \tab{tab2}.) Note, that both outermost columns are identical. They have been duplicated to facilitate easier comparison.}
      \label{fig:lder}
   \end{center}
\end{figure}

\section{Results}

\begin{figure}[t]
   \includegraphics[width=\columnwidth]{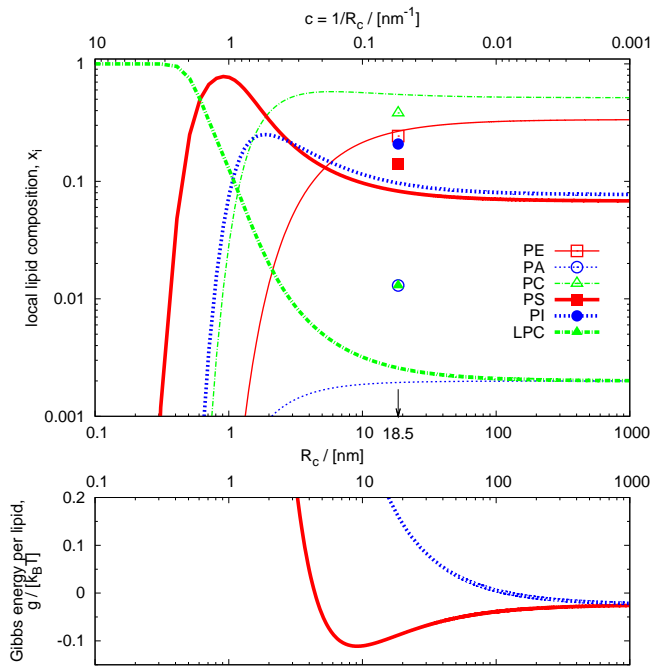}
   \caption{Predicted lipid composition, $x_i (\rc)$, of a PL monolayer covering a spherical surface of radius $\rc$ (lines in upper panel). For $\rc\rightarrow\infty$, the curves slowly converge to the measured composition $\xpi$ of the planar ER membrane \cite{zinser1991}. The lipid composition of ripe, detached LD, as determined by \cite{zinser1991}, are indicated by symbols at the position of best fit (arrow). Lower panel, average free Gibbs energy per lipid, $g$ as function of cap radius, $\rc$.  It shows the impact of spherical bending on $g$ for a compositionally optimized lipid membrane (full line) and a compositionally rigid membrane with frozen-in ER composition (dashed line).}
   \label{fig:ldcomp}
\end{figure}

The optimal PL composition, $x_i$, of a LD membrane was evaluated as
function of cap radius, $\rc$, using
\eqn{molefraction}. \fig{fig:ldcomp} illustrates the predicted lipid
distribution in the LD monolayer (upper panel) originating from a
planar ER with PL composition according to \cite{zinser1991}. $\rc$
represents the radius of a spherical protuberance in an otherwise
planar lipid layer (\fig{fig:ldcapneck}). Its inverse, $c=1/\rc$, may
be interpreted as the mean curvature of the associated sphere and
\fig{fig:ldcomp} can be understood in terms of the curvature
dependence of the lipid composition.

\fig{fig:ldcomp} describes the demixing process that occurs for curved
membranes. For small curvatures ($c<0.005$~nm$^{-1}
\Leftrightarrow\rc>200$~nm), the local PL composition in the ER
membrane is only weakly affected and close to the lipid distribution
of the unperturbed planar ER. On the other hand, for small cap radii
($\rc < 100$~nm $\Leftrightarrow c>0.01$~nm$^{-1}$ ), a significant
demixing is observable; while generally lipids with $S_i>1$ (PE, PA,
and PC; thin lines in \fig{fig:ldcomp} ) decrease with decreasing
$\rc$, the fractions of PS, PI, and LPC ($S_i<1$; thick lines)
considerable increase. The observation is consistent with
expectation, as lipids with $S > 1$ tend to form convex surfaces,
while $S < 1$ favors concave topologies. However, this trend changes
at high curvatures, where PI and PS content peak. Thus, for even
higher curvatures, i.e. $\rc < 1$ nm, the monolayer enriches only on
LPC at the expense of all other PL.

PC is an exception to the rule stated above. Despite its $S$ value being larger than 1 ($S_i=1.11$) it does not migrate but slightly increases for intermediate values of $\rc$. Eventually, however, even PC drifts off for $\rc < 5$ nm. Only a small fraction of this discrepancy can be attributed to the differences in elastic moduli. For instance, if $\kci$ is changed from $9\kbt$ to $10\kbt$ (plus the corresponding changes in $\kcbari$), then the maximum PC content in the membrane decreases by only $2\%$ (data not shown).

In the lower panel of \fig{fig:ldcomp} we show the change of Gibbs energy per lipid accompanied by the demixing process (full line) according to \eqn{gcap}. For comparison we have also plotted the $\rc$ dependence for a membrane with frozen-in ER composition [dotted line; \eqn{gcap} together with the setting $x_i=\xpi, \forall\rc$]. Both lines converge for $\rc\rightarrow\infty$. However, while for a membrane with fixed composition the energy increases with decreasing $\rc$, the behavior is more subtle for a membrane, which optimizes its lipid composition. In this case the Gibbs energy initially decreases with decreasing $\rc$, reaching a pronounced minimum at $\rc \approx 9$ nm. Thus an optimized packing is able to overcompensate the increase in elastic energy caused by spherical distortion of an originally planar surface. Only at very high curvatures is regrouping of lipids unable to balance the rise in bending energy and the Gibbs energy sharply increases.

\begin{figure}[t]
   \includegraphics[width=\columnwidth]{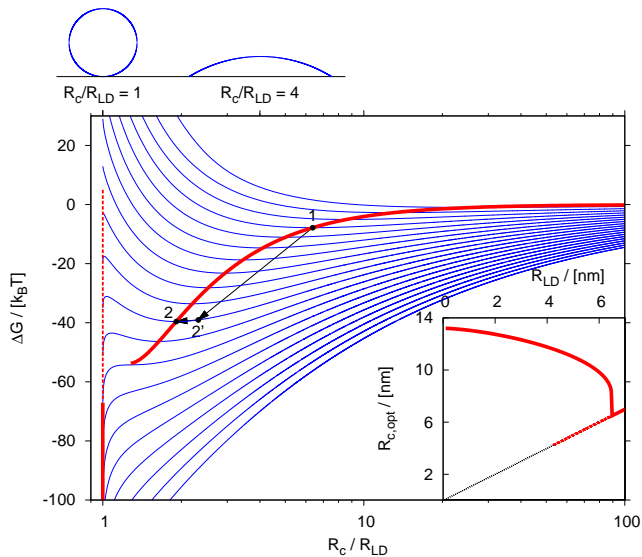}
   \caption{Total free Gibbs energy, $\dg(\rc;\rld)$, as function of the normalized cap radius, $\rc/\rld$. Each full, thin line represents the energy dependence for a fixed but constant LD volume ($\rld=[0.5,10]$ nm, increment $0.5$ nm). The observed (local) minima are marked by thick lines. For a better understanding of the meaning of the normalized $x$-axis the effective geometrical conformation is illustrated for two exemplary values on top. $\rc$ denotes the cap radius, while $\rld$ represents the equivalent spherical radius of the cap. Inset: Position of the local energy minima, $\rcopt$, as function of LD size, $\rld$. The thick lines correspond to the thick lines in the main figure. The dotted line indicate points where $\rcopt=\rld$.}
   \label{fig:totdg}
\end{figure}
To predict the total energy of formation for a nascent LD of volume $4\pi R^3_{\text{LD}}/3$ and cap radius $\rc$ we evaluate \eqn{eqn:totdg}. \fig{fig:totdg} shows $\dg(\rc;\rld)$ as function of $\rc$. Note that $\dg$ depends parametrically on the LD-volume, represented by $\rld$. That is, each line in \fig{fig:totdg} shows the dependency of the energy on $\rc$ for a fixed LD volume. Depending on the size of the LD, that is on the value of $\rld$, the free Gibbs energy shows either a single minimum at values $\rc/\rld>1$ (for $\rld<4.2$ nm), two minima -- one at $\rc/\rld=1$ and the other at $\rc/\rld>1$ (for $4.2$ nm $<\rld<6.5$ nm) --, or exactly one minimum at $\rc/\rld=1$ (for $\rld>6.5$ nm).

\fig{fig:totdg} may be interpreted in such a way that for constant LD volume, minima in the Gibbs energy single out preferred configurations, i.e. identify likely values of $\rc$ for a given LD volume (thick full and dashed line). $\rc/\rld=1$ indicates that a LD does no longer form a spherical cap, but a full sphere. Thus, if the minimum in energy shows up at such a point, the LD has been fully fabricated and detaches from the ER. If, however, a minimum appears at values $\rc/\rld>1$, then the nascent LD forms a stable spherical protuberance of volume $4\pi R^3_{\text{LD}}/3$ and cap radius $\rc$, but remains an integral part of the ER membrane. The position of the minima, $\rcopt$, as function of LD size, $\rld$, is illustrated in the inset of \fig{fig:totdg}. In this plot, points along the dotted line ($\rc/\rld=1$) indicate fully detached LD.

The stability of a nascent LD retained in the ER may be best estimated by the height of the energy barrier between the position of the minimum, $\rcopt$, and the point where the cap becomes a full sphere, i.e. at $\rc=\rld$.  This barrier shrinks with increasing LD volume. In our calculation we find that for LD diameters larger than $2\rld= 13$ nm this barrier vanish completely and does not reappear again. Thus such LD are fully matured and detach from the ER. It is important to note that if we keep the lipid composition of the ER membrane fixed, then no decreasing energy barrier is observable (data not shown), and LD remain as integral part of the ER -- in contrast to experimental findings. Thus lipid demixing is an essential part in the budding process. Moreover, we conclude that our lipid demixing model can explain the budding process based on biophysical reasons only. 

To understand the reason for the disappearance of the energy barrier
consider the change in energy associated with LD formation,
\eqn{eqn:totdg}. In essence $\dg(\rc;\rld)$ depends on the product
between $\gcap(\rc)$ and $\ac(\rc;\rld)$. Since $\ac(\rc;\rld)$ has
values just for $\rc\ge\rld$, only these distances contribute to the
product. On the other hand, $\gcap(\rc)$ does not depend on $\rld$,
and neither does its minimum at $\rc \approx 9$ nm. Therefore, if
$\rld$ is small enough so that $\rld$ lies to the left of the minimum
in \fig{fig:ldcomp}, then also the total energy exhibits a pronounced
minimum and $\dg(\rc;\rld)$ rises sharply in-between. An energy
barrier is formed. On the other hand, if $\rld$ lies to the right of
the minimum, then $\gcap(\rc)$ is cut-off before it reaches its
minimum, a stabilizing energy barrier cannot form and LD detach from
the ER. Thus the position of the minimum in $\gcap(\rc)$ in relation
to the LD size, i.e. $\rld$, is decisive for the appearance of an
energy barrier due to $\gcap(\rc)$.

The predicted bud-off diameter depends on the initial ER membrane composition. To what extent the bud-off radius can be controlled by the ER membrane composition has not been investigated. However, we successfully checked whether we get similar results if the PL composition of LD rather than the one of the ER is used as an input for the simulation. Therefore we solved \eqn{molefraction} such that at a typical LD size of $\rc=200$ nm \cite{czabany2008} the measured PL composition \cite{zinser1991} is retrieved. In that case we predicted a bud-off diameter of $6.7$ nm (see supplementary material \ref{sec:LDdata}, \fig{fig:ldcompLD} and \fig{fig:totdgLD}).

Using the parameter settings in \tab{tab1}, our model is able to describe budding. At the bud-off diameter, $2\rld=13$ nm, we observed a significantly altered membrane composition in comparison with the ER. Consistently, we got a similar scale for the budding size, if we fit $\rc$ such that the computed composition (taken from \fig{fig:ldcomp}) best approximates the measured data. By doing so, we obtained $\rld=18.5$ nm and $\rld=3.2$ nm for the data in \cite{zinser1991} and \cite{leber1994}, respectively (see \fig{fig:lipstruc}). However, in any case our predicted bud-off radius is an order of magnitude lower than the typical LD radius of $200$ nm \cite{czabany2008}. This suggested that if the standard model of LD biogenesis is correct, then it is a two-step process. Step one, growth of a nano-sized LD nucleus within the ER membrane. Step 2, cytosolic ripening of a nano-LD after they have budded-off from the ER.

\section{Discussion}

\paragraph*{Lipid demixing in curved membrane regions.}
We have presented a thermodynamic model to describe LD formation and investigated the interplay between membrane curvature and demixing in the PL composition. In fact, lipids favoring a convex surface ($S<1$, i.e. PS, PI, LPC) are enriched while PE, PA, and PC are depleted. Curvature induced demixing is hardly surprising and already well documented \cite{roux2005}. All the more important, however, is the observation that the free energy per lipid for a
curvature-optimized packing can drop below that of a planar monolayer. In other words, moderately deforming an originally planar mixed PL monolayer does not necessarily cost energy, but can result in an energy gain -- provided that the lipid system can optimize its composition.

\paragraph*{Demixing forces induce membrane curvature.}
Our analysis showed that during lipid demixing of membranes energy is gained rather than consumed. Thus we argue that demixing at least supports if not drives membrane deformation. Such a view is evoked by \fig{fig:totdg}. By increasing the NL content in the ER one moves between two isolines (transition from 1 to 2'; note that due to normalization the transition appears as a tilted line even though $\rc$ remained constant). At 2' the membrane perceives a "demixing force" which further distorts the monolayer and pushes the system toward a new minimum at point 2, thereby sucking NL in the nascent LD. Note that in \fig{fig:totdg} demixing forces occur to the right of the thick line, while to its left forces are fueled by elastic energy. Thus demixing is a general feature of our model. It does not require LD biogenesis proceeding along the energy minima in \fig{fig:totdg}.

At this point it has become clear that only the existence of a spherical surface -- as a topological alternative to the plane -- renders the molecular shape an energetically relevant quantity. This makes it energetically worthwhile for the membrane to separate the different lipids according to their molecular shapes. In that sense, the spherical topology is the real agent of the predicted demixing process.

\paragraph*{Demixing -- a potential method to recruit proteins from the ER membrane?}
We speculate that not only lipids but also tracer components of the ER membrane are enriched or depleted on nascent LD as dictated by their specific molecular shape factor. In principle, such a mechanism would be able to explain the clustering of specific proteins on raised membrane domains. For example, cone-shaped membrane proteins (as long as they are not transmembrane proteins), should react to the positive curvature of developing LD. This mechanism is potentially able to explain the so far unaccounted accumulation of over-expressed caveolin-1 on LD \cite[check cermelli2006]{ostermeyer2001,cermelli2006}. Caveolin-1 has a hairpin-like structure, with both terminal facing the cytosol. Thus, its structure factor is smaller than 1 and therefore, like PS, enriches on the LD surface. Likewise, proteins with molecular shapes akin to PE should be depleted. This offers the interesting perspective of an efficient, self-organized, biophysical ``protein sorting'' mechanism that performs the recruiting of proteins from the pool of ER proteins according to their molecular shape.

\paragraph*{The birth of a lipid droplet.}
We argued that PL demixing in developing LD is not a consequence of the membrane curvature, but rather its driving force. The existence of pronounced minima in the Gibbs energy (\fig{fig:totdg}) leads to demixing forces during LD growth. Thus the following thermodynamic, budding mechanism is suggested: A LD forms by accumulating NL within the ER membrane, thereby producing a membrane protuberance. Its curvature is adjusted by a demixing force in such a way that it minimizes the interfacial energy of the cytoplasmic monolayer (see \fig{fig:totdg}, transition 2' to 2). As long as the NL volume, i.e. the size of the protuberance measured by $\rld$, is small enough, the LD is prevented from detaching by an insurmountable energy barrier. This barrier is traced back to the sharp rise in the elastic energy at very high curvatures. However, for NL volumes, $V$, for which $\rld=\sqrt[3]{3 V/(4 \pi)} > 6.5$ nm, the developing LD does not buckle enough. The increase in bending energy is too small and a stabilizing energy barrier cannot mold. Thus, such large LD do not exist within the ER membrane, as they have budded off already at an earlier stage of their development.

At first it might appear counterintuitive that the elastic properties of a thin skin such as a PL monolayer should ultimately be responsible for the formation of LD. However, considering a LD suspension as some sort of macroemulsion, these ideas are far from new: in fact, it has been shown  that type and stability of oil-in-water emulsions are determined by sign and value of the monolayer's spontaneous curvature \cite{kabalnov1996}.

LD size distributions based on electron microscopic pictures for
various yeast mutant strains, are well reproducible and rather narrow,
showing values between 250 nm and 550 nm in diameter, with a maximum
at about 400 nm \cite{czabany2008}. Nevertheless these radii are an
order of magnitude larger than those predicted by our model. What
could be the reason for this discrepancy? Obviously, the
experimentally observed distribution reflects the size of the LD
suspension in the cytosol. This distribution does not necessarily have
to be identical to the size of freshly released LD; implying that
there exists a second, subsequent phase of LD ripening within the
cytosol. Thus, small LD produced from the ER would undergo a series of
fusions through conglomeration and coalescence until they reach their
full-grown size \cite{murphy1999}. A number of observations support
this view.

Although current experimental methods allow for the detection of 400 nm structures, a fully matured LD, which is still part of the ER, has never been observed in any cell. Neither have smaller ones. This is probably because such LD are too small to be resolved with conventional experimental techniques \cite{lacey1999,fujimoto2008}. Even with the usage of electron microscopy, which provides the ultimate resolution, nascent LD would be difficult to observe. By cutting a section of 80-90 nm through a fixed cell it is unlikely to section also an attached LD. Nevertheless, very recently large NL globules were detected by electron microscopy within special sub-compartments of the ER membrane in human hepatoma cells \cite{ohsaki2008}. The authors hypothesized that tight binding of anomaly lipidated apolipoproteins suppressed detaching and allowed detecting giant NL structures within the ER membrane. Whether this observation presents the missing link in validating the standard model of LD formations remains to be seen.

TAG accumulates either in LD dispersed in the cytosol or in membrane
micro-domains \cite{hakumaki2000}. These micro-domains are tiny TAG
depots residing between two membrane leaflets
\cite{mountford1988}. Such depots were estimated to have a size
between 22-28 nm which is roughly consistent with our diameters.  This
would suggest that these small inclusions are simply nascent
LD. However, based on their data \cite{hakumaki2000}, an ultimate
conclusion is not possible.

Artificially prepared PL membranes on a solid surface showed
time-dependent changes in the membrane topography after externally
initiating NL production \cite{waltermann2005}. Droplet-like
structures were formed at the surface with an average diameter of
about 50 nm. Again, this is comparable to the size regime suggested by
our model. Then, in a second process stage, smaller LD coalesced into
larger structures. Additionally, LD fusion has also been shown \emph{in
vivo} \cite{bostrom2005}.

If LD biogenesis proceeds in a two step process, i.e. budding of nano-LD, which later merge, then fusion will dramatically change the surface to volume ratio and we also expect alterations in the PL composition of the LD, too \cite{kuerschner2008}. In fact, if, for simplicity, we assume that LD bud-off at a diameter of 40 nm then $1,000$ nano-LD are needed to produce a typically sized LD. The total surface of these nano-LD will be ten times larger than the surface of the full-blown LD. This excess of PL may be used as a reservoir, which would allow the growing LD to re-mix its PL composition in order to accommodate the decreasing curvature. However, this reservoir will not be accessible as a whole, as it is unlikely that all $1,000$ nano-LD merge simultaneously in a single big fusion. Thus lipid re-mixing due to fusing of LD ultimately depends on the process of coalescence. Nevertheless we expect it to be less efficient then the original lipid demixing in the ER membrane. In fact, in our model we obtained PL compositions for stable nano-LD that are similar to those of fully matured ones (see \fig{fig:lder}). Thus, the lack of accessible PL-depots during the ripening process could explain why nano-LD and matured LD have similar compositions, despite their different sizes. 

However, any remixing would change the composition but leaves the total number of excess PL unaltered. On the other hand, many LD are found to be surrounded by or adherent to double layer structures  \cite[(unpublished, H Wolinski)]{kuerschner2008}. These membranes are neither part of the ER nor do they form continuous structures with LD. Their origin and function is unclear. Here we speculate that these membranes are a waste dump sprouting from excess PL during LD fusion.

It is also interesting to link our results to mammalian very low-density lipoproteins (VLDL), which are another kind of lipid particles. Their NL core are of the same order of magnitude ($30 - 90$ nm) as our predicted nano-LD \cite{ohsaki2009}. This observation would support the so far untested hypothesis of \cite{alexander1976} who suggested that ready-made nano-LD fuse with apoproteins to form VLDL.

\paragraph*{Model limitations.}
Our model assumes that LD, though forming between the leaflets of the
ER membrane, lead to a protuberance only of the cytoplasmic leaflet,
while the endoplasmic leaflet remains unaffected. Such an asymmetry
may be real and caused by certain membrane proteins, differences in
the PL composition of the leaflets or different properties of the
solvents adjacent to the leaflets. However, a rigid endoplasmic
leaflet is not an essential assumption. Our model could be adopted to
account for LD growth in both leaflets.

An essential part of our approach is the spherical cap approximation, which has been proven helpful in the context of the Young equation, which relates the contact angle of a liquid droplet on a solid surface to the inter-facial energies of all three phases involved \cite{butt2003}. However, this approximation neglects contribution from the neck region. The neck connects the planar ER membrane with the LD monolayer and has the geometrical shape of a torus. In contrast to the cap, the neck region consist of both, positive and negative curvatures. 

Our current model is not able to address neck formation. However, it was demonstrated theoretically that the neck shape is generated by an interplay between the local curvature and the membrane composition \cite{iglic2007,haegerstrand2006}. Similar to the observation reported here (accumulation of $S<1$-lipids at high curvatures), it was shown that the neck region predominately consists of lipids with anisotropic, saddle-like intrinsic curvature. This is consistent with our results as these lipids are best adapted to the torical shaped neck. We are currently working on including the neck in our description. However, we do not expect any significant impact on our conclusions, because firstly the neck region is small in comparison to the spherical cap and secondly contributions to the total energy will be attenuated by lipid demixing.

\section{Summary}
A biophysical model to describe LD biogenesis has been presented. We have estimated the local composition of a PL monolayer, which is part of a spherical protuberance in an otherwise planar membrane. Four major results are obtained. (i) In order to use the available surface area economically, PL with shape factors, $S < 1$ (PI, PS, LPC), tend to accumulate on spherical surfaces. Conversely, lipids with $S > 1$ (PE, PC, PA) migrate. Therefore, a spherical protuberance of radius, $\rc$, in an otherwise planar membrane causes a local lipid demixing on the curved membrane monolayer. (ii) As a consequence of the optimized packing of lipids, the energy per lipid $\gcap(\rc)$ drops with decreasing $\rc$, i.e., with increasing curvature. Thus, induced by lipid demixing, a monolayer gains energy by forming a spherical protuberance of weak to intermediate curvature. Only at very high curvatures, $\gcap(\rc)$ increases again, leading to a clear minimum in $\gcap(\rc)$ at $9$ nm. (iii) This energy minimum is key in understanding the formation of an energy barrier, which controls the budding of LD. (iv) The height of the energy barrier depends on the volume of the protuberance; it stabilizes protuberances of smaller volumes, but completely vanishes at higher volumes. Our model predicts that LD detach from the ER at a diameter in the order of $13$ nm.

In this paper we have suggested that LD formation is driven by lipid demixing. This finding is based on a model calculation. The main ingredient entering this model is the geometrical structure of various lipids, described by the shape factor. Thus, by changing the shape factor, we expect different demixing effects and ultimately different bud-off sizes, which could be verified by \emph{in vitro} measurements, similar to those performed by \cite{waltermann2005}. Experimentally the shape factor may be influenced by varying the chain length of FA in lipids, or by manipulating the pH of the solvent which influences the head group areas.

\paragraph*{Acknowledgement} Intensive discussions with Sepp D. Kohlwein, G\"unther Daum, Martin Peifer, Georg Pabst, Mihnea Hristea, and Klaus Natter are gratefully acknowledged.

This work was supported by a grant from the Austrian Federal Ministry for Science and Research (Project GOLD within the framework of the Austrian GEN-AU program)

\bibliographystyle{apsper}
\bibliography{ldbio}

\clearpage
\appendix
\section{Estimating structural parameters of PL}
\label{sec:lipest}
PE is the only lipid for which both, $\aoi=54$ \AA$^2$ and $\boi=105$ \AA$^2$ have been determined experimentally \cite{fuller2001}. Together with $\coi=-0.043$ \AA$^{-1}$ \cite{rand1990} and \eqn{scurv}, one obtains $l=22$ \AA, which is roughly half the thickness of a typical PE bilayer ranging between 40 and 50 \AA~\cite{rappolt2004}. It also compares well with the value $l=18$ \AA~obtained by \cite{marsh1996}.

To estimate data for PC, we followed an idea of Marsh \cite{marsh2007}, who estimated $\co$ values for lipid dioleoylphosphatidyl-PE/DOPC mixtures by fitting experimental data \cite{rand1990}. We assumed $\aoi$ of PC to be 72 \AA$^2$ \cite{marsh1996,nagle2000} and used \eqn{scurv} as fitting function for the data in \cite{rand1990}. An excellent fit is obtained, giving a value of $\boi=87.8$ \AA$^2$ and subsequently $\coi=-0.01$ \AA$^{-1}$, which agrees with the values given in \cite{marsh1996,chen1997,szule2002}.

For PA, PI, and PS the situation is less clear. As we have been unable to find experimentally measured head-group areas of PA and PI, we have estimated these values by extrapolating the linear relationship between molecular volume and head-group area obtained for PC, PE and O-LPC. Also, the molecular area at the pivotal plane $\api$ are unknown for PS, PI and PA; again, values have been guessed by comparison with corresponding values of other lipids.
\begin{table}[b]
\caption{\label{tab2} Literature values for experimentally (E) measured PL composition of ER ($\xpi$) and LD ($x_i$) membranes compared to the results of our thermodynamic model (C). Data listed in \cite{zinser1991,leber1994} for ``other lipids'' have been redistributed such that each column sums up to one. $\rc$ values denote the radii of best fit for the experimental data. Calculated PL compositions represent values at the radii of best fit.}
\begin{ruledtabular}
\begin{tabular}{lllllll}
                   & $\xpi$(E\footnotemark[1]) & $x_i$(E\footnotemark[1]) &
  $x_i$(C) & $x_i$(E\footnotemark[2])  & $x_i$(C)                              \\
  \hline
 PE              & 0.336 & 0.243 & 0.266 & 0.203 & 0.059 \\
 PC              & 0.515 & 0.382 & 0.549 & 0.368 & 0.537 \\
 PS              & 0.068 & 0.141 & 0.083 & 0.055 & 0.192 \\
 PI                & 0.077 & 0.208 & 0.097 & 0.32   & 0.202 \\
 O-LPC       & 0.002 & 0.013 & 0.003 & 0.027 & 0.009 \\
 PA              & 0.002 & 0.013 & 0.002 & 0.027 & 0.001 \medskip\\
 
 $\rc$ (nm) & $\infty$ & 18.5 & 18.5   & 3.2      & 3.2    \\
\end{tabular}
\end{ruledtabular}
\begin{minipage}{0.17\columnwidth}
\footnotetext[1]{Ref. \cite{zinser1991}}
\end{minipage}\hfill
\begin{minipage}{0.17\columnwidth}
\footnotetext[2]{Ref. \cite{leber1994}}
\end{minipage}\hfill
\begin{minipage}{0.17\columnwidth}
\quad
\end{minipage}\hfill
\begin{minipage}{0.17\columnwidth}
\quad
\end{minipage}\hfill
\begin{minipage}{0.17\columnwidth}
\quad
\end{minipage}

\end{table}

\clearpage

\section{Changing the PL composition of the ER membrane}
\label{sec:LDdata}

\begin{figure}[!h]
   \includegraphics[width=\columnwidth]{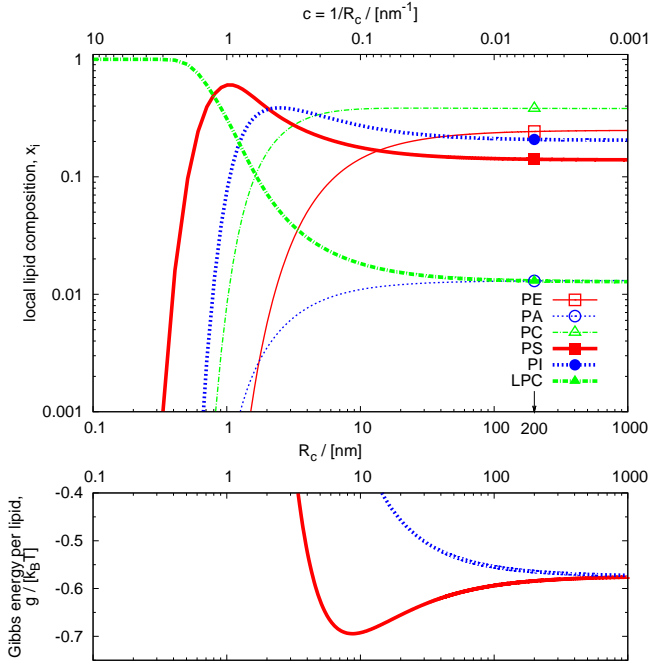}
   \caption{Predicted lipid composition, $x_i (\rc)$, of a PL monolayer covering a spherical surface of radius $\rc$ (lines in upper panel). The PL composition of the ER membrane was adjusted such that at $\rc= 200$ nm (arrow) the measured PL composition of a LD is met \cite{zinser1991}.}
   \label{fig:ldcompLD}
\end{figure}

\begin{figure}[t]
   \includegraphics[width=\columnwidth]{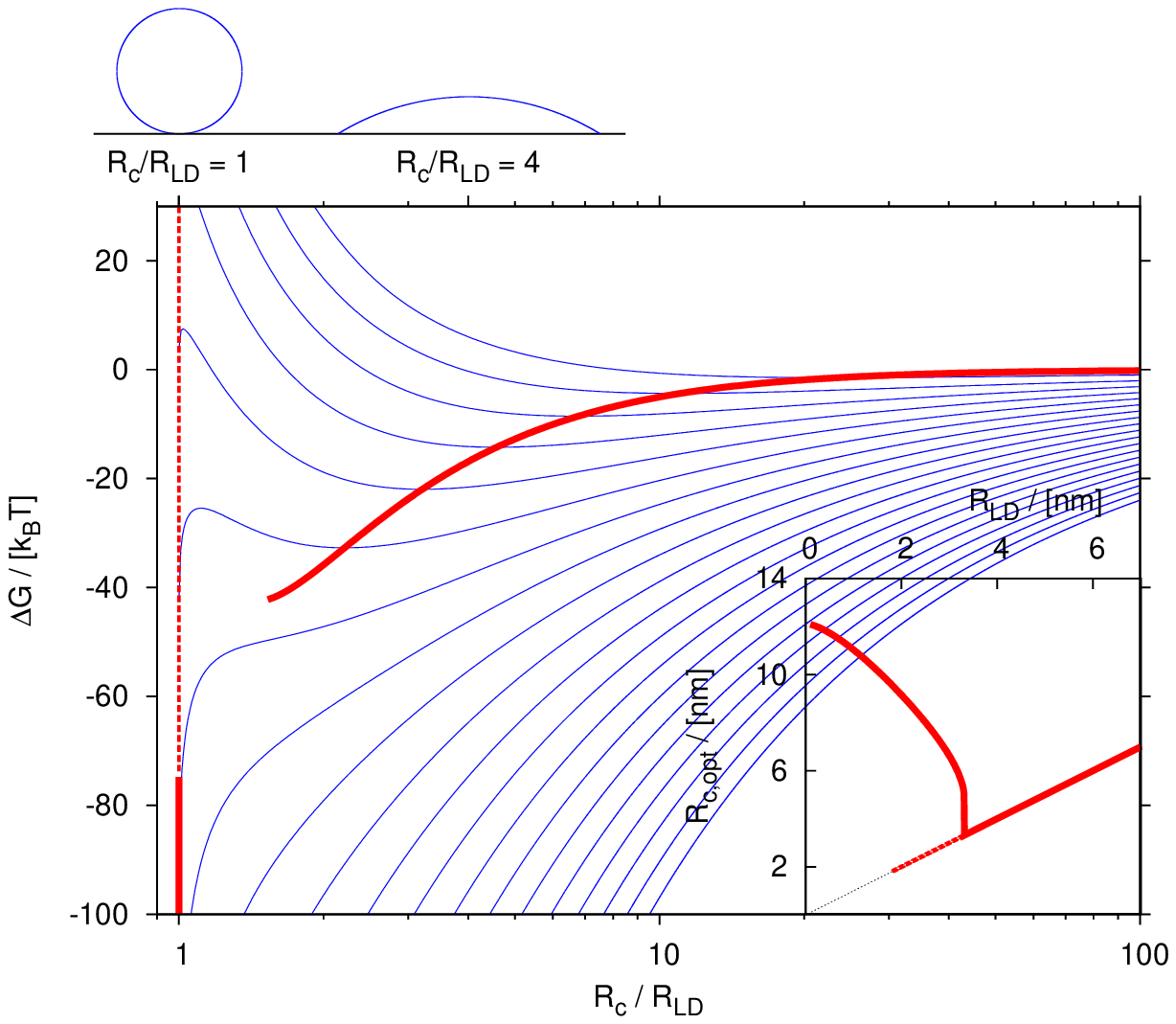}
   \caption{Total free Gibbs energy, $\dg(\rc;\rld)$, -- for the data of \fig{fig:ldcompLD} -- as function of the normalized cap radius, $\rc/\rld$. Each full, thin line represents the energy dependence for a fixed but constant LD volume ($\rld=[0.5,10]$ nm, increment $0.5$ nm). The observed (local) minima are marked by thick lines. For a better understanding of the meaning of the normalized $x$-axis the effective geometrical conformation is illustrated for two exemplary values on top. Inset: Position of the local energy minima, $\rcopt$, as function of LD size, $\rld$. The thick lines correspond to the thick lines in the main figure. The dotted line indicate points where $\rcopt=\rld$.}
   \label{fig:totdgLD}
\end{figure}

\begin{table}[b]
\caption{\label{tab3} Experimentally (E) and computationally (C) obtained PL composition for ER ($\xpi$) and LD ($x_i$) membranes. The calculated PL composition of the ER membrane was adjusted such that at $\rc= 200$ nm the measured PL composition of a LD is met \cite{zinser1991}. Data listed in \cite{zinser1991} for ``other lipids'' have been redistributed such that each column sums up to one. $\delta_rx=[\xpi(C)-x]/x$ denotes the relative error with respect to $x$.}
\begin{ruledtabular}
\begin{tabular}{ldddddd}
                   & \multicolumn{1}{r}{$x_i$(E)} & \multicolumn{1}{r}{$\xpi$(E)} &
  \multicolumn{1}{r}{$\xpi$(C)} & \multicolumn{1}{r}{$\delta_r\xpi$} & \multicolumn{1}{r}{$\delta_rx_i$}          \\
  \hline
 PE              & 0.243 & 0.336 & 0.248 & -0.262 & 0.021 \\
 PC              & 0.382 & 0.515 & 0.381 & -0.260 & -0.003 \\
 PS              & 0.141 & 0.068 & 0.140 &  1.059 & -0.007 \\
 PI                & 0.208 & 0.077 & 0.205 &  1.662 & -0.014 \\
 O-LPC       & 0.013 & 0.002 & 0.013 & 5.5 & 0.000 \\
 PA              & 0.013 & 0.002 & 0.013 & 5.5 & 0.000 \\[1.5ex]
 $\rc$ (nm) & 200 &  \multicolumn{1}{c}{$\infty$} &  \multicolumn{1}{c}{$\infty$} &&\\
\end{tabular}
\end{ruledtabular}

\end{table}

\clearpage

\end{document}